\documentstyle[12pt]{article}
\begin{document}
\begin{center}
{\large EXTENDED SPACE DUALITY IN THE NONCOMMUTATIVE PLANE   }\\
\vskip 2cm
Subir Ghosh\\
\vskip 1cm
Physics and Applied Mathematics Unit,\\
Indian Statistical Institute,\\
203 B. T. Road, Calcutta 700108, \\
India.
\end{center}
\vskip 3cm
{\bf Abstract:}\\
Non-Commutative (NC) effects  in planar quantum mechanics are investigated.  We have constructed a {\it{Master}}
model for a noncommutative harmonic oscillator by embedding it in an
extended space, following the Batalin-Tyutin \cite{bt} prescription.
Different gauge choices lead to distinct NC structures, such as NC
coordinates, NC momenta  or noncommutativity of a more general kind.
In the present framework, all of these
can be studied in a unified and systematic manner. Thus the dual
nature of theories having different forms of noncommutativity
is also revealed.


\vskip 2cm
\noindent {\bf{Introduction:}} Non-Commutative (NC) Quantum Field
Theories (QFT) \cite{rev} have created a lot of interest in the
High Energy Physics community because of its direct connection to
certain low energy limits of String theory \cite {sw}.
Noncommutativity is induced in the open string boundaries that are
attached to D-branes, in the presence of a two-form background
field. This phenomenon in turn renders the D-branes into NC
manifolds. Stringy effects are manifested in NCQFT framework. The
major advantage of working in the latter is that the basic
structure of QFT in conventional (commutative) spacetime remains
intact and the fundamental (two-point) correlation functions of
QFT are not modified by NC effects. This vital fact emerges from
the construction of the NC generalization of a conventional QFT
where
the products of the field operators in the QFT action are replaced by $*$
-product (or Moyal-Weyl product) defined below,
\begin{equation}
\hat f(x)*\hat g(x)=e^{\frac{i}{2}\theta_{\mu\nu}\partial_{\sigma_{\mu}}%
\partial_{\xi_{\nu}}}\hat f(x+\sigma )\hat g(x+\xi )\mid_{\sigma =\xi =0}
=\hat f(x)\hat g(x)+\frac{i}{2}\theta^{\rho\sigma}\partial_{\rho}\hat
f(x)\partial_{\sigma}\hat g(x)+~O(\theta^{2}),  \label{1}
\end{equation}
where $\theta_{\mu\nu}$ is conventionally taken as a constant anti-symmetric
tensor. $\hat f(x)$ stands for the NC extension of $f(x)$. {\footnote{$f(x)$ and $\hat f(x)$
 are related in a non-trivial way by the Seiberg-Witten Map when quantum
gauge theories are being considered. However, this is of no concern in the
present work.}}

Exploiting the above rule (\ref{1}), one derives the spacetime
noncommutativity in $\hat x_\mu $-space,
\begin{equation}
[\hat x_{\mu}, \hat x_{\nu}]_{*}= \hat x_\mu * \hat x_\nu - \hat x_\nu *
\hat x_\mu =i\theta_{\mu\nu},  \label{2}
\end{equation}
where $\hat x_\mu $ is the NC (operator) analogue of $x_\mu $.
In the present work we will only be concerned with spatial noncommutativity
in
a plane and so (\ref{2}) reduces to
\begin{equation}
\{x^i,x^j\}=\theta\epsilon^{ij}~;~~\epsilon^{12}=1.  \label{2a}
\end{equation}
Note that in our classical setup we will interpret the noncommutativity as
a modification in the symplectic structure.

Effects of noncommutativity are often analyzed in terms of
momentum variables \cite{rev}. This is because  in the conventional
forms of NC theory, the momentum
degrees of freedom  behave in a canonical way, obeying
the symplectic structure,
\begin{equation}
\{\hat p_{i},\hat p_{j}\}=0~;~~\{\hat x_{i},\hat p_{j}\} =\delta_{ij}.  \label{40}
\end{equation}
Here the coordinates are noncommutative given by (\ref{2a}).

In case of NC quantum field theories, NC effects modify only the
interaction
vertices (through momentum dependent phase factors)  when the fields
are expressed by their Fourier transforms in terms of
momentum variables.

On the other hand, in generic NC quantum mechanical models, one
replaces \cite{nair,mitra} the original coordinates by a canonical set of
variables carrying a representation of the NC algebra. In terms of this new
set of variables, the NC effects are manifestly present in the Hamiltonian.
As a concrete example, the NC phase space algebra (\ref{2a},\ref{40}) can
be simulated \cite{nair,mitra} by the canonical variables $(Q_i,P_j)$
obeying
$$\{Q_{i},P_{j}\}=\delta_{ij},~\{Q_{i},Q_{j}\}=\{P_{i},P_{j}\}=0,$$
with the identification,
 \begin{equation}
x_{i}\equiv Q_{i}-\frac{\theta}{2} \epsilon _{ij}P_{j}~;~~p_{i}\equiv P_{i}.  \label{3b}
\end{equation}

The NC extension of the ordinary oscillator Hamiltonian is,
\begin{equation}
\hat H=\frac{1}{2m}\hat p_{i}*\hat p_{i}+\frac{1}{2}\hat x_{i}*\hat x_{i}=\frac{1}{2m}\hat p_{i}\hat p_{i}+\frac{1}{2}\hat x_{i}\hat x_{i},~~i=1,2  \label{3a}
\end{equation}
where we have used the following relations,
$$
\hat x_{i}\hat x_{i}=\hat x_{i}\hat x_{i}+\frac{i}{2}\theta \epsilon _{ij}\delta
_{ij}\equiv~(\hat x)^2;~~\hat p_{i}*\hat p_{i}=\hat p_{i}\hat p_{i}\equiv
(\hat p)^{2}.
$$
NC effects will become manifest in $H$ once the $(\hat x,\hat p)$
variables are replaced by $(Q,P)$ using (\ref{3b}).

 Let us now define the perspective of our work. Essentially, we have
 generalized the above $(P,Q)$-model of NC harmonic oscillator to
 a {\it {Master }} model by embedding it in an extended phase space,
 in the Batalin-Tyutin (BT) framework \cite{bt}. The advantages of
 working with the {\it{Master}} model are the following: It can generate
 different structures of noncommutativity and ensures at the same time that
 the latter are gauge equivalent, (to be explained below). In contrast,
 note that the $(Q,P)$-system, as defined in (\ref{3b}) is geared to
 induce only the particular form of noncommutativity (\ref{2a}-\ref{40}).
 However, keeping in mind subsequent (canonical) quantization of the model,
 we wish to maintain the canonical structure of the phase space of the
 new variables and BT formulation \cite{bt} is tailor made  for that purpose.

\vskip 1cm
\noindent {\bf{BT prescription:}} Let us digress briefly on the BT
formalism \cite{bt}. The basic idea is to embed the original system
in an enlarged phase space (the BT space), consisting of the original
"physical" degrees of freedom and auxiliary variables, in a particular
way such that the resulting enlarged system possesses local gauge invariance. Imposition of gauge conditions accounts for the true number of degrees of freedom and at the same time the freedom of having different gauge choices leads to structurally distinct systems. However, all of them are assured to be gauge equivalent.  In fact, this method is a generalization of the well known Stuckelberg formalism \cite{stu}, which is applicable in the Lagrangian framework.

In the Hamiltonian analysis of constrained systems, as formulated by
Dirac \cite{dirac}, the constraints are termed
as First Class Constraints (FCC) if they commute (in the Poisson
Bracket (PB) sense,
 modulo constraints) or Second Class Constraints (SCC)
if they do not. The FCCs induce gauge invariance in the theory
whereas the SCCs tend to modify the symplectic structure of the
phase space for compatibility with the SCCs. The above modification
induces a replacement of
the PBs by Dirac Brackets (DB)  as defined below,
\begin{equation}
\{A,B\}_{DB}= \{A,B\}-\{A,\psi _\alpha \}
(\Psi _{\alpha\beta})^{-1}\{\psi _\beta ,B\}.
\label{dirac}
\end{equation}
where $\psi _\alpha $ refer to the SCCs and $\Psi _{\alpha \beta }\equiv \{\psi _{\alpha },\psi _{\beta }\}$ is invertible on the constraint surface. The Dirac brackets are compatible with the SCCs so that the SCCs can be put "strongly" to zero.

 Let us consider a generic set of constraints $(\psi _\alpha ,\xi _l
)$ and a Hamiltonian operator $H$ with the following PB relations,
$$
\{\psi _\alpha  (q) ,\psi _\beta (q)\}\approx
\Delta _{\alpha \beta
 }(q) \ne 0~~;~~\{\psi _\alpha (q) ,\xi _l (q)\}\approx 0
$$
\begin{equation}
 \{\xi _l(q) ,\xi _n (q)\}\approx 0 ~~;~~
\{\xi _l(q) ,H (q)\}\approx 0.
\label{01}
\end{equation}
In the above $(q)$
collectively refers to the set of variables
present prior to the BT extension
and "$\approx $"
means that the equality holds on the constraint surface. Clearly
$\psi
_\alpha  $
and $\xi _l $ are SCC and FCC  respectively.

In systems with non-linear SCCs,
(such as the present one), in general the DBs can become
dynamical variable dependent
due to the $\{A,\psi _\alpha \}$ and
$\Delta _{\alpha \beta }$
 terms, leading to problems for
the quantization programme. This type of pathology is cured in the BT
formalism in a systematic way, where one introduces the BT
variables $\phi
 ^\alpha  $, obeying
\begin{equation}
\{\phi ^\alpha ,\phi ^\beta \}=\omega ^{\alpha \beta}=
-\omega ^{\beta \alpha},
\label{bt}
\end{equation}
where $\omega ^{\alpha \beta}$ is a constant (or at most
 a c-number function) matrix, with the aim of modifying the SCC
$\psi _\alpha  (q)$ to $\tilde \psi _\alpha  (q,
\phi ^\alpha )$
such that,
\begin{equation}
\{\tilde\psi _\alpha (q,\phi ) ,\tilde\psi _\beta (q,\phi )\}=0
~~;~~\tilde\psi _\alpha (q,\phi )=\psi _\alpha (q)+
\Sigma _{n=1}^\infty \tilde\psi ^{(n)} _\alpha (q,\phi )~~;~~
\tilde\psi ^{(n)}\approx O(\phi ^n)
\label{b1}
\end{equation}
This means that $\tilde\psi  _\alpha $ are now FCCs and in
particular abelian.
The explicit terms in the above expansion are,
\begin{equation}
\tilde\psi ^{(1)}_\alpha =X_{\alpha \beta }\phi ^\beta ~~;~~
\Delta _{\alpha \beta }+X_{\alpha \gamma }
\omega ^{\gamma \delta }X_{\beta \delta }=0
\label{b2}
\end{equation}

\begin{equation}
\tilde\psi ^{(n+1)}_\alpha =-{1\over{n+2}}
\phi^{\delta }\omega _{\delta \gamma }X^{\gamma \beta }B^{(n)}_{\beta \alpha }~~;~~n\ge 1
\label{b3}
\end{equation}

\begin{equation}
B^{(1)}_{\beta \alpha }=
\{\tilde\psi ^{(0)} _\beta ,\tilde\psi ^{(1)} _\alpha \}_{(q)}-
\{\tilde\psi ^{(0)} _\alpha ,\tilde\psi ^{u(1)} _\beta \}_{(q)}
\label{b4}
\end{equation}

\begin{equation}
B^{(n)}_{\beta \alpha }=
\Sigma _{m=0}^n
\{\tilde\psi ^{(n-m)} _\beta ,\tilde\psi ^{(m)} _\alpha \}_{(q,p)}+
\Sigma _{m=0}^n
\{\tilde\psi ^{(n-m)} _\beta ,\tilde\psi ^{(m+2)} _\alpha \}_{(\phi )}
~~;~~n\ge 2
\label{b5}
\end{equation}
In the above, we have defined,
\begin{equation}
X_{\alpha \beta }X^{\beta \gamma }=
\omega _{\alpha \beta }\omega ^{\beta \gamma }
=\delta ^\gamma _\alpha \delta  .
\label{b6}
\end{equation}
A very useful
idea is to introduce the improved function $\tilde f(q)$
 corresponding to each $f(q)$,
\begin{equation}
\tilde f(q,\phi )\equiv f(\tilde q)
=f(q)+\Sigma _{n=1}^\infty \tilde f(q,\phi )^{(n)}~~
;~~\tilde f^{(1)}=-
\phi^{\beta }\omega _{\beta \gamma }X^{\gamma \delta }\{
\psi_\delta ,f\}_{(q)}
\label{b7}
\end{equation}

\begin{equation}
\tilde f^{(n+1)}=-{1\over{n+1}}
\phi^{\beta }\omega _{\beta \gamma }X
^{\gamma \delta }
G(f)^{\lambda (n)}_\delta ~~;~~n\ge 1
\label{b8}
\end{equation}

\begin{equation}
G(f)^{(n)}_{\beta }=
\Sigma _{m=0}^n
\{\tilde\psi ^{ (n-m)} _\beta ,\tilde f^{(m)}\}_{(q)}+
\Sigma _{m=0}^{(n-2)}
\{\tilde\psi ^{ (n-m)} _\beta ,\tilde f^{(m+2)}\}_{(\phi )}
+\{\tilde\psi ^{ (n+1)} _\beta ,\tilde f^{(1)}\}_{(\phi )}
\label{b9}
\end{equation}
which have the property
$\{\tilde\psi _\alpha (q,\phi ) ,\tilde f(q,\phi )\}=0$.
Thus, in the BT space, the improved functions are FC or equivalently
gauge invariant. Note that $\tilde q$ corresponds to the improved
variables for $q$.
The subscript $(\phi )$ and $(q)$ in the PBs indicate the  variables
with respect to which the PBs are to be taken.
It can be
proved that extensions of the original FCC $\xi _l $ and Hamiltonian
 $H$ are simply,
\begin{equation}
\tilde \xi _l=\xi (\tilde q)~~;~~
\tilde H=H (\tilde q).
\label{b10}
\end{equation}
One can also reexpress the converted SCCs as
$\tilde\psi ^\mu_\alpha \equiv \psi ^\mu_\alpha (\tilde q)$.
The following identification theorem holds,
\begin{equation}
\{\tilde A,\tilde B \}=\tilde {\{A,B\}_{DB}}~~;~~
\{\tilde A,\tilde B \}\mid _{\phi =0}=\{A,B \}
 _{DB}~~;~~\tilde 0=0.
\label{b11}
\end{equation}
Hence the outcome
of the BT extension is the closed system of FCCs with the FC
Hamiltonian given below,
\begin{equation}
\{\tilde \psi _\alpha  ,\tilde \psi _\beta \}=
\{\tilde \psi _\alpha  ,\tilde \xi _l\}=
\{\tilde \psi _\alpha ^\mu ,\tilde H\}= 0~~;~~
\{\tilde \xi _l ,\tilde \xi _n\}\approx 0 ~;~
\{\tilde \xi _l ,\tilde H\}\approx 0.
\label{b12}
\end{equation}
In general, due to the non-linearity in the SCCs, the extensions
 in the improved variables, (and subsequently in the FCCs
 and FC Hamiltonian), may
turn out to be infinite series. This type of situation has been encountered
 before \cite{sg}. However, this is not the case in the present work.

\vskip 1cm
\noindent
{\bf{NC harmonic oscillator:}}  We will work with a specific
model \cite{rb} for an NC harmonic oscillator,
\begin{equation}
L=q_{i}\dot{x}_{i}+\frac{\theta }{2}\epsilon _{ij}q_{i}\dot{q}_{j}-\frac{k}{2
}x^{2}-\frac{1}{2m}q^{2}.  \label{10}
\end{equation}
The connection between $L$ in (\ref{10}) and the conventional models of
noncommutativity arising
in the coordinates of charged particle moving in a plane with a normal
magnetic field is
discussed in \cite{rb}. This first order Lagrangian  possesses the
following set of constraints,
\begin{equation}
\psi _{1}^{i}\equiv \pi _{x}^{i}-q^{i}~;~~\psi _{2}^{i}\equiv \pi _{q}^{i}+
\frac{\theta }{2}\epsilon ^{ij}q^{j}  \label{11}
\end{equation}
and the non-singular nature of the commutator matrix for the constraints
\begin{equation}
\Psi _{\alpha \beta }^{ij}=\{\psi _{\alpha }^{i},\psi _{\beta }^{j}\}~;~~\alpha ,\beta \equiv 1,2
\label{11a}
\end{equation}
where
$$
\Psi _{\alpha \beta }^{ij}=\left(
\begin{array}{cc}
0 & -\delta ^{ij} \\
\delta ^{ij} & \theta \epsilon ^{ij}
\end{array}
\right)
$$
indicates that the constraints are Second Class constraints (SCC) in the
Dirac terminology \cite{dirac}  The
SCCs require a change in the symplectic structure in the form of Dirac
Brackets \cite{dirac} defined in (\ref{dirac}).
 In the present case, the
inverse of the constraint matrix (\ref{11a})
$$
\Psi _{\alpha \beta }^{(-1)ij}=\left(
\begin{array}{cc}
\theta \epsilon ^{ij} & \delta ^{ij} \\
-\delta ^{ij} & 0
\end{array}
\right)
$$
leads to the following set of Dirac brackets,
$$
\{x^{i},x^{j}\}_{DB}=\theta \epsilon ^{ij}~,~~\{x^{i},q^{j}\}_{DB}=\delta
^{ij}~;~~\{q^{i},q^{j}\}_{DB}=0,  $$
\begin{equation}
\{q^{i},\pi_q^{j}\}_{DB}=0~;~~\{x^{i},\pi_x^{j}\}_{DB}=\delta
^{ij}~;~~\{\pi_x^{i},\pi_q^{j}\}_{DB}=0.
\label{11c}
\end{equation}
The Hamiltonian is
\begin{equation}
H=\frac{k}{2}x^{2}+\frac{1}{2m}q^{2}.  \label{12}
\end{equation}
Note the spatial noncommutativity and also the fact that $q^i$
behaves effectively as the conjugate momentum to $x^i$. Although,
explicit $\theta $-dependence does not show up in the Hamiltonian,
it will appear in the Hamiltonian equations of motion where the
Dirac brackets  are to be used. The model does not have any FCC
in the physical space.

\vskip 1cm
\noindent
{\bf{BT extension of NC harmonic oscillator:}}
Let us now turn to the main body of our work.
  It is important to remember that the extended space
is {\it{completely}} canonical and the Dirac
brackets of (\ref{11c}) are not to be used. This is
because the SCCs (\ref{11}) are absent in the extended
space and their place is taken by the FCCs that we derive below.

Following the procedure outlined in the previous section, we
introduce a canonical set of auxiliary variables
\begin{equation}
\{\phi _{\alpha }^{i},\phi _{\beta }^{j}\}=\epsilon _{\alpha \beta }\delta
^{ij}~,\alpha ,\beta =1,2~;~~\epsilon _{12}=1,  \label{13}
\end{equation}
it is possible to convert the SCCs in (\ref{11}) to
the following FCCs $\tilde{\psi}_{\alpha }^{i}$,
$$
\tilde{\psi}_{1}^{i}\equiv \psi _{1}^{i}+\phi _{2}^{i}=\pi
_{x}^{i}-q^{i}+\phi _{2}^{i},
$$
\begin{equation}
\tilde{\psi}_{2}^{i}\equiv \psi _{2}^{i}-\phi _{1}^{i}-\frac{\theta }{2}%
\epsilon ^{ij}\phi _{2}^{j}
=\pi _{q}^{i}+\frac{\theta }{2}\epsilon ^{ij}q^{j}-\phi _{1}^{i}-\frac{%
\theta }{2}\epsilon ^{ij}\phi _{2}^{j} , \label{14}
\end{equation}
so the $\tilde \psi ^i_\alpha $ are commutating,
\begin{equation}
\{\tilde{\psi}_{\alpha }^{i},\tilde{\psi}_{\beta }^{j}\}=0.  \label{15}
\end{equation}
Thus the embedded model in extended space possesses local gauge
invariance. Let us construct the improved variables $\tilde q$ as
defined in (\ref{b7}-\ref{b9}).
 The connection
between the operator algebra in any reduced ({\it{i.e.}} gauge fixed)
offspring with its analogue  in the
gauge invariant parent model is the following identity:
\begin{equation}
\{A,B\}_{DB}=\{\tilde A, \tilde B\}.
\label{id}
\end{equation}
In the present case we compute FC counterparts of $x^i$ and $q^i$:
$$
\tilde{q}^{i}=q^{i}-\phi _{2}^{i}~,~~\tilde{x}^{i}=x^{i}-\phi _{1}^{i}+\frac{
\theta }{2}\epsilon ^{ij}\phi _{j}^{2},$$
\begin{equation}
\tilde \pi _x^i =\pi _x^i ~,~~\tilde \pi _q^i =\pi _q^i-\phi^{i}_{1},
  \label{17}
\end{equation}
which in turn generates the FC Hamiltonian
\begin{equation}
\tilde{H}=\frac{k}{2}\tilde{x}^{2}+\frac{1}{2m}\tilde{q}^{2}
=\frac{k}{2}(x^{i}-\phi _{1}^{i}+\frac{\theta }{2}\epsilon ^{ij}\phi
_{2}^{j})^{2}+\frac{1}{2m}(q^{i}-\phi _{2}^{i})^{2}.  \label{18}
\end{equation}
The FCCs take a simpler form,
\begin{equation}
\tilde{\psi}_{1}^{i}= \tilde\pi
_{x}^{i}-\tilde q^{i}~;~~\tilde{\psi}_{2}^{i}
=\tilde\pi _{q}^{i}+\frac{\theta }{2}\epsilon ^{ij}\tilde q^{j}.
\label{ffc}
\end{equation}
This is the cherished form of the Hamiltonian where the NC correction has
appeared explicitly in a fully canonical space with commuting space
coordinates. However we would like to keep the set of dynamical equations
\cite{rb}
\begin{equation}
\dot{q}^{i}=-kx^{i}~;~~\dot{x}^{i}=\frac{1}{m}q^{i}+\theta k\epsilon
^{ij}x^{j}  \label{19}
\end{equation}
unchanged and this can be achieved by adding suitable terms in the
Hamiltonian that are
proportional to the FCCs.  We construct $\tilde
H_{Total}$
\begin{equation}
\tilde{H}_{Total}=\tilde{H}+\lambda _{1}^{i}\tilde\psi _{1}^{i}+\lambda
_{2}^{i}\tilde\psi _{2}^{i}  \label{20}
\end{equation}
and identify the arbitrary multipliers $\lambda ^i_\alpha $ from (\ref{20}):
\begin{equation}
\dot{q}^{i}=\{q^{i},\tilde{H}_{Total}\}=\lambda _{2}^{i}~;~~\dot{x}%
^{i}=\{x^{i},\tilde{H}_{Total}\}=\lambda _{1}^{i}.  \label{21}
\end{equation}
Hence the final form of the gauge invariant FC Hamiltonian generating the
correct dynamics of \cite{rb} is
$$
\tilde{H}_{Total}=\frac{k}{2}(x^{i}-\phi _{1}^{i}+\frac{\theta }{2}\epsilon
^{ij}\phi ^{j}_{2})^{2}+\frac{1}{2m}(q^{i}-\phi _{2}^{i})^{2}
$$
\begin{equation}
+(\frac{1}{m}q^{i}+\theta k\epsilon ^{ij}x^{j})(\pi _{x}^{i}-q^{i}+\phi
_{2}^{i})-kx^{i}(\pi _{q}^{i}+\frac{\theta }{2}\epsilon ^{ij}q^{j}-\phi
_{1}^{i}-\frac{\theta }{2}\epsilon ^{ij}\phi _{2}^{j}).  \label{22}
\end{equation}
This  is the {\it{Master}} model that we advertised in the Introduction
and constitute the main result of the present paper. The local gauge
invariance in the enlarged space allows us to choose gauge conditions
(according to our convenience) which in turn gives rise to different
forms of
symplectic structures and Hamiltonians, that , however, are gauge
{\it{equivalent}}. This is the duality between different structures of
noncommutativity,
referred to earlier. As an obvious gauge choice, the unitary gauge
\begin{equation}
\psi _{3}^{i}\equiv \phi _{1}^{i}~,~~\psi _{4}^{i}\equiv \phi
_{2}^{i}  \label{23}
\end{equation}
restricts the system to the original physical subspace with the spatial
noncommutativity as given in (\ref{11c}) being induced
by the full set of SCCs (the  FCCs $\tilde \psi ^i_1 ,\tilde \psi ^i_2$
and the
gauge fixing constraints $\psi ^i_3,\psi ^i_4$).

On the other hand, the following non-trivial gauge
\begin{equation}
\psi _{3}^{i}\equiv \phi _{2}^{i}~;~~\psi _{4}^{i}\equiv \phi _{1}^{i}-\frac{%
\theta }{2}\epsilon ^{ij}\phi _{2}^{j}+c\epsilon ^{ij}\pi _{x}^{j},
\label{24}
\end{equation}
with $c$ being an arbitrary parameter, generates the constraint matrix
\begin{equation}
\Psi _{\alpha \beta }^{ij}=\{\tilde{\psi}_{\alpha }^{i},\psi _{\beta }^{j}\}
\label{25}
\end{equation}
where,
$$
\Psi _{\alpha \beta }^{ij}=\left(
\begin{array}{cccc}
0 & 0 & 0 & -\delta ^{ij} \\
0 & 0 & -\delta ^{ij} & 0 \\
0 & \delta ^{ij} & 0 & -\delta ^{ij} \\
\delta ^{ij} & 0 & \delta ^{ij} & \theta \epsilon ^{ij}
\end{array}
\right) .
$$
The inverse matrix
$$
\Psi _{\alpha \beta }^{(-1)ij}=\left(
\begin{array}{cccc}
\theta \epsilon ^{ij} & \delta ^{ij} & 0 & \delta ^{ij} \\
-\delta ^{ij} & 0 & \delta ^{ij} & 0 \\
0 & -\delta ^{ij} & 0 & 0 \\
-\delta ^{ij} & 0 & 0 & 0
\end{array}
\right)
$$
induces the Dirac brackets,
\begin{equation}
\{x^{i},x^{j}\}_{DB}=(\theta +2c)\epsilon ^{ij}~;~~\{x^{i},\pi _{x}^{j}\}_{DB}=\delta
^{ij}~;~~\{\pi _{x}^{i},\pi _{x}^{j}\}_{DB}=0.  \label{26}
\end{equation}
The choice $c=-\theta /2$ reduces the
phase space to a canonical one with  the Hamiltonian
\begin{equation}
H=\frac{k}{2}x^{2}+(\frac{1}{2m}+\frac{k\theta ^{2}}{8})\pi _{x}^{2}-\frac{
k\theta }{2}\epsilon ^{ij}x^{i}\pi _{x}^{j}.  \label{27}
\end{equation}
Rest of the non-trivial Dirac brackets are not important in the present
case.
This structure is in fact identical to the one studied in \cite{mitra}.

Another interesting gauge is
\begin{equation}
\psi _{3}^{i}\equiv q^{i}-\phi _{2}^{i}+ax^{i}~;~~\psi _{4}^{i}\equiv
x^{i}-\phi _{1}^{i}+\frac{\theta }{2}\epsilon ^{ij}\phi _{2}^{j}+bq^{i},
\label{27a}
\end{equation}
where $a$ and $b$ are arbitrary parameters.
Once again, the subsequent constraint matrix
$$
\Psi _{\alpha \beta }^{ij}=\left(
\begin{array}{cccc}
0 & 0 & -a\delta ^{ij} & 0 \\
0 & 0 & 0 & -b\delta ^{ij} \\
a\delta ^{ij} & 0 & 0 & -\delta ^{ij} \\
0 & b\delta ^{ij} & \delta ^{ij} & \theta \epsilon ^{ij}
\end{array}
\right)
$$
and its inverse
$$
\Psi _{\alpha \beta }^{(-1)ij}=\left(
\begin{array}{cccc}
0 & -\frac{\delta ^{ij}}{ab} & \frac{\delta ^{ij}}{a} & 0 \\
\frac{\delta ^{ij}}{ab} & \frac{\theta \epsilon ^{ij}}{b^{2}} & 0 & \frac{%
\delta ^{ij}}{b} \\
-\frac{\delta ^{ij}}{a} & 0 & 0 & 0 \\
0 & -\frac{\delta ^{ij}}{b} & 0 & 0
\end{array}
\right)
$$
results in the symplectic structure,
\begin{equation}
\{x^{i},x^{j}\}_{DB}=0~;~~\{q^{i},q^{j}\}_{DB}=\frac{\theta }{b^{2}}\epsilon
^{ij}~;~~\{x^{i},q^{j}\}_{DB}=-\frac{\delta ^{ij}}{ab}.  \label{29}
\end{equation}
The choice of the parameters $a=\pm 1,b=\mp 1$ fixes $q^i$ to be the
conjugate momentum to $x^i$  but {\it{the momentum variables have
now become noncommutative}}. Comparing the Hamiltonian
\begin{equation}
H=\frac{k}{2}q^{2}+\frac{1}{2m}x^{2}  \label{30}
\end{equation}
with (\ref{12}) (where $x^i$ were noncommutative) one observes that
$q^i$ and $p^i$ have replaced one another. This exercise clearly
demonstrates the fact that coordinate or momentum noncommutativity
are actually different sides of the same coin and are connected by
gauge transformations.

We conclude with a brief comment on the angular momentum $L$ of the system.
Remembering that in the extended space, the symplectic structure is
completely canonical, $L$ will have the obvious form
\begin{equation}
L=\epsilon^{ij}(x^i\pi^{j}_{x}+q^i\pi^{j}_{q}+\phi^{i}_{1}\phi^{j}_{2}).
\label{l1}
\end{equation}
Upon utilizing the canonical commutators, $L$ will generate
correct transformations on the degrees of freedom. But notice that
$L$ is not gauge invariant ({\it{i.e.}} FC) in the extended space.
Its FC generalization will be,
\begin{equation}
L=\epsilon^{ij}(\tilde x^i\tilde \pi^{j}_{x}+\tilde q^i\tilde \pi^{j}_{q}),
\label{l2}
\end{equation}
since $\tilde\phi^{i}_{\alpha}=0$. To recover the correct transformations of the physical variables one has to construct $L_{Total}$,
$$L_{Total}=L+\xi^{i}_{\alpha}\tilde \psi^{i}_{\alpha},$$
\begin{equation}
\xi^{i}_{1}=\epsilon^{ij}(-\phi^{j}_{1}+\frac{\theta}{2}\epsilon^{jk}\phi^{k}_{2})~;~~\xi^{i}_{2}=-\epsilon^{ij}\phi^{j}_{2}.
\label{l3}
\end{equation}
In the unitary gauge $\tilde L_{Total}$ reduces to the expression given in \cite{rb}. For other gauge choices the structure of angular momentum will be different and the corresponding Dirac brackets will reproduce the transformation.

\vskip 1cm
{\bf{Conclusion:}} We have generalized the noncommutative harmonic oscillaor to a {\it{Master}} model having gauge invariance, in a Batalin-Tyutin extended space Hamiltonian framework. The NC effects manifest themselves
through the embedding procedure. We have shown that different gauge choices lead to different Hamiltonian systems with  distinct
symplectic structures.
This clearly demonstrates the duality (or gauge equivalence) between
different types of noncommutativity, (such as between coordinates,
between momenta and between both coordinates as well as momenta), that are
induced by different gauge choices in the same {\it{Master}} model.

\end{document}